\documentclass[aps,twocolumn,prx]{revtex4-2}

\usepackage{xcolor,hyperref}
\hypersetup{
   colorlinks,
   linkcolor={blue!50!black},
   citecolor={blue!50!black},
   urlcolor={blue!80!black}
}

\usepackage{epsfig, amsmath, graphicx}
\usepackage{bm}
\usepackage{color}

\usepackage[latin1]{inputenc}
\newcommand{\beq}{\begin{equation}}
\newcommand{\eeq}{\end{equation}}
\newcommand{\bea}{\begin{eqnarray}}
\newcommand{\eea}{\end{eqnarray}}
\newcommand{\pa}{\partial}

\renewcommand{\=}{\!=\!}

\begin{document}

\title{Self-healing (solitonic) slip pulses in frictional systems}
\author{Anna Pomyalov$^{1}$}
\thanks{A. Pomyalov and Y. Lubomirsky contributed equally.}
\author{Yuri Lubomirsky$^{1}$}
\thanks{A. Pomyalov and Y. Lubomirsky contributed equally.}
\author{Lara Braverman$^{1,2}$}
\author{Efim A.~Brener$^{3,4}$}
\author{Eran Bouchbinder$^{1}$}
\email{eran.bouchbinder@weizmann.ac.il}
\affiliation{$^{1}$Chemical and Biological Physics Department, Weizmann Institute of Science, Rehovot 7610001, Israel\\
$^{2}$Department of Physics, The University of Chicago, Chicago, Illinois 60637, USA \\
$^{3}$Peter Gr\"unberg Institut, Forschungszentrum J\"ulich, D-52425 J\"ulich, Germany\\
$^{4}$Institute for Energy and Climate Research, Forschungszentrum J\"ulich, D-52425 J\"ulich, Germany}

\begin{abstract}
A prominent spatiotemporal failure mode of frictional systems is self-healing slip pulses, which are propagating solitonic structures that feature a characteristic length. Here, we numerically derive a family of steady state slip pulse solutions along generic and realistic rate-and-state dependent frictional interfaces, separating large deformable bodies in contact. Such nonlinear interfaces feature a non-monotonic frictional strength as a function of the slip velocity, with a local minimum. The solutions exhibit a diverging length and strongly inertial propagation velocities, when the driving stress approaches the frictional strength characterizing the local minimum from above, and change their character when it is away from it. An approximate scaling theory quantitatively explains these observations. The derived pulse solutions also exhibit significant spatially-extended dissipation in excess of the edge-localized dissipation (the effective fracture energy) and an unconventional edge singularity. The relevance of our findings for available observations is discussed.
\end{abstract}
\maketitle

{\em Introduction}.---Understanding the emergence and properties of compact spatiotemporal structures in driven dissipative systems, featuring long-range interactions, is of prime importance in a wide variety of physical systems~\cite{newell1985solitons,dauxois2006physics}. A prominent example is frictional systems, typically composed of two deformable bodies in contact along a frictional interface, e.g.~a geological fault in the earth's crust. Upon the application of external driving forces, slippage at the interface commences, accompanied by partial rupture of the frictional contact and frictional strength reduction. This failure process --- e.g.~a propagating earthquake --- is intrinsically inhomogeneous and often takes the form of spatially-compact, solitonic slip pulses~\cite{Freund1979,Heaton1990,Perrin1995,Beroza_Mikumo_1996,Beeler1996,Cochard1996,Andrews1997,Zheng1998,Nielsen2000,Nielsen2003,Brener2005,Shi2008,Rubin2009,Garagash2012,Gabriel2012,Putelat2017,Michel2017,Brener2018,brantut2019stability,lambert2021propagation,ROCH2022104607}.

Slip pulses are self healing in nature~\cite{Perrin1995}. They feature significant strength reduction near their leading edge that invades a nearly quiescent, non-slipping interfacial state, but also strength recovery at their trailing edge, involving interfacial healing. Consequently, they feature a characteristic slipping length $L$ over which nearly stationary contact is recovered. The healing/restrengthening process at the pulse trailing edge is intimately related to the nonequilibrium nature of frictional interfaces, which are known to undergo contact aging under nominally quiescent conditions~\cite{Dieterich1978,Dieterich1994a,Beeler1994,Marone1998a,Berthoud1999,Baumberger2006Solid,Ben-David2010}. In between the two edges, slip pulses feature finite slip velocities $v$. Numerous geophysical observations, laboratory experiments and numerical simulations demonstrated that driven frictional systems can spontaneously generate long-lived, self-healing slip pulses~\cite{Freund1979,Heaton1990,Perrin1995,Beroza_Mikumo_1996,Beeler1996,Cochard1996,Andrews1997,Zheng1998,Nielsen2000,Nielsen2003,Brener2005,Shi2008,Rubin2009,Garagash2012,Gabriel2012,Putelat2017,Michel2017,Brener2018,brantut2019stability,lambert2021propagation,ROCH2022104607}. Yet, understanding the existence and properties of such self-healing slip pulses remains incomplete and challenging.

The problem involves a coarse-grained interfacial constitutive law, which relates the frictional strength $\tau(v,\ldots)$ to the slip velocity $v$ and to a set of internal state fields represented by the ellipsis~\cite{Dieterich1979a,Ruina1983,Marone1998a,Nakatani2001,Baumberger2006Solid}. The latter describe the structural state of the interface at each spatial point at any time, corresponding to an evolving ensemble of contact asperities, and play the role of nonequilibrium order parameters. $\tau(v,\ldots)$ is intrinsically nonlinear and accounts for significant energy dissipation. The nonlinear and dissipative interfacial constitutive law is coupled to the elastodynamic deformation of the bodies forming the interface, implying that distant parts of the interface are coupled by long-range spatiotemporal elastic forces. Finally, frictional systems are typically driven externally by far-field forces, e.g.~a homogeneous shear stress $\tau_{\rm d}$ applied at the boundaries of the bodies in contact.

Considering two large and identical linear elastic bodies in frictional contact, described by a spatial coordinate $x$, and focusing on objects propagating steadily at a velocity $c_{\rm p}$, the interplay between the various physical ingredients discussed above is encapsulated in the following nonlinear integral equation~\cite{Weertman1980}
\begin{equation}
\tau[v(\xi),\phi(\xi)]=\tau_{\rm d}-\mu\,{\cal F}(\beta)\!\int_{-\infty}^{\infty}\frac{v(z)}{z-\xi}dz
\label{eq:f_xi} \ ,
\end{equation}
where $\xi\=x-c_{\rm p}t$ in a co-moving coordinate ($t$ is time). $\tau(v,\phi)$ is the frictional strength that depends, in addition to the slip velocity $v$, also on the field $\phi(\xi)$ that quantifies the amount of interfacial contact. The dynamics of the internal state field $\phi$ account for the competition between contact aging and slip-induced rejuvenation, to be discussed below.

$\tau(v,\phi)$ equals the interfacial shear stress, corresponding to the right hand side of Eq.~\eqref{eq:f_xi}. It is composed of the external driving shear stress $\tau_{\rm d}$ and of a weighted integral over the slip velocity field $v(\xi)$, which represents the long-range elastodynamic interaction between different parts of the interface. $\mu$ is the shear modulus of the bodies forming the frictional interface and ${\cal F}(\beta)$ is a known function that accounts for material inertia, where $\beta\!\equiv\!c_{\rm p}/c_{\rm s}$ is the dimensionless propagation velocity and $c_{\rm s}$ is the shear wave-speed of the bodies.

Steady state pulse solutions to Eq.~\eqref{eq:f_xi} are those that satisfy the self-healing boundary conditions $v(\xi\!\to\!\pm\infty)\!\to\!0$, i.e.~solutions that feature a spatially-compact slip velocity field $v(\xi)$. In this work, we numerically derive a family of such solutions for a generic and physically realistic interfacial constitutive law $\tau(v,\phi)$~\cite{Dieterich1979a,Ruina1983,Marone1998a,Nakatani2001,Baumberger2006Solid,Bar-Sinai2014}. We thoroughly analyze the properties of the emerging slip pulses and theoretically explain them. Finally, the relevance of our findings for available observations are briefly discussed.

{\em The interfacial constitutive law}.---Eq.~\eqref{eq:f_xi}, together with the self-healing boundary conditions, constitute a well defined problem once $\tau(v,\phi)$ and the evolution equation for $\phi(\xi)$ are specified. Over the last few decades, it has been established that the local contact area at frictional interfaces, which is typically orders of magnitude smaller than the nominal contact area, grows with the stationary contact ($v\!\to\!0$) time $t$ in proportional to $\log(1+t/\phi_*)$, where $\phi_*$ is the contact aging onset time~\cite{Dieterich1978,Dieterich1994a,Beeler1994,Marone1998a,Berthoud1999,Baumberger2006Solid,Ben-David2010}. This contact aging leads to frictional strengthening and is described by an internal state field $\phi$ of time dimension, which identifies with $t$ for $v\!\to\!0$. Conversely, under steady sliding conditions at a slip velocity $v$, $\phi$ has been shown to be proportional to $1/v$~\cite{Ben-David2010}, leading to contact rejuvenation and frictional weakening. The transition between the aging and steady sliding regimes occurs over a characteristic slip distance $D$. These observations are accounted for by~\cite{Marone1998a,Nakatani2001,Baumberger2006Solid}
\begin{equation}
-\beta\,c_{\rm s}\,\partial_{\xi}\phi(\xi)= 1-\frac{|v(\xi)|\phi(\xi)}{D}
\label{eq:dxi_phi}\ ,
\end{equation}
in terms of the co-moving coordinate $\xi$.

\begin{figure}[ht!]
\centering
\includegraphics[width=0.5\textwidth]{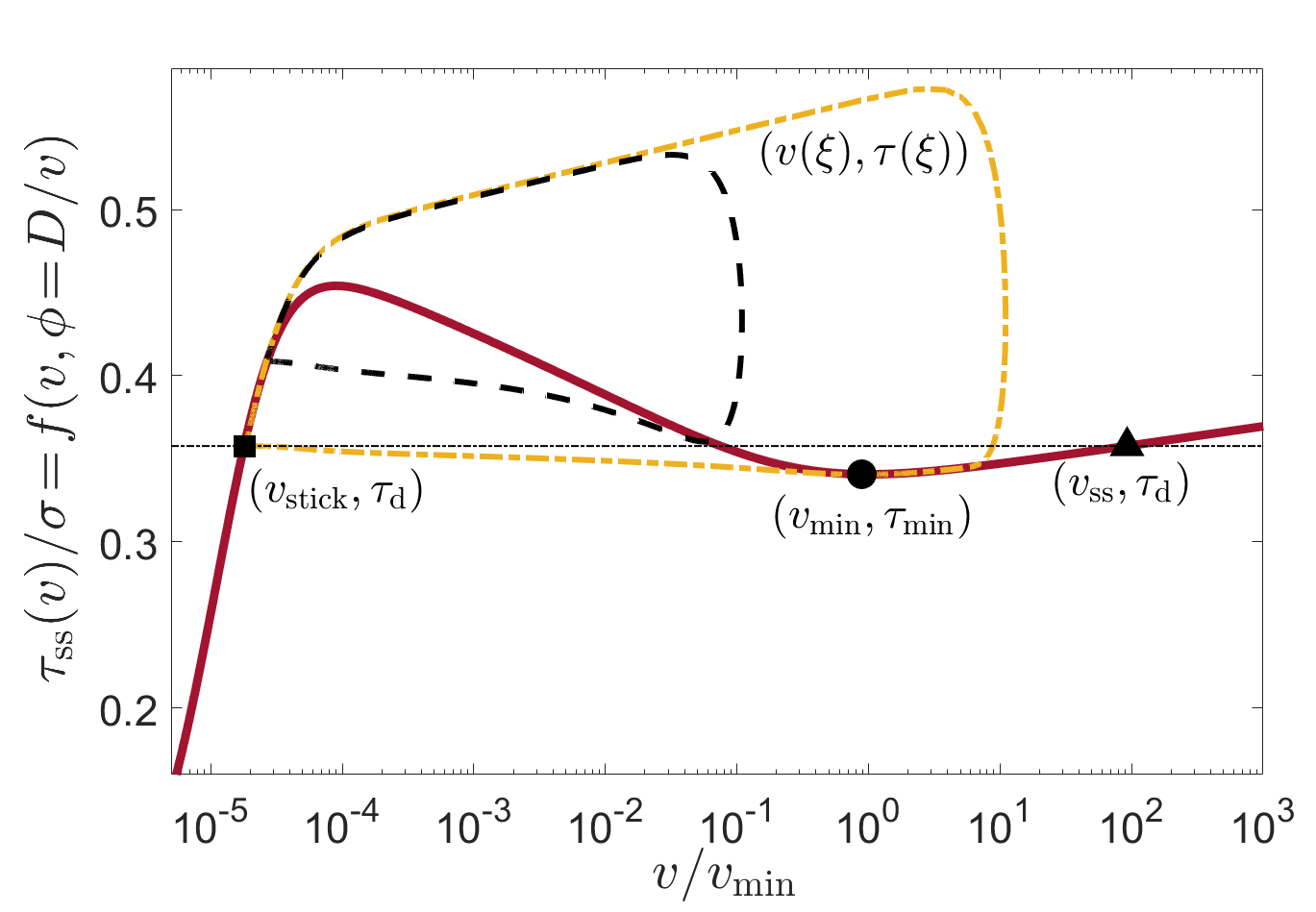}
\caption{$\tau_{\rm ss}$ (solid line, in units of the normal stress $\sigma$) vs.~$v/v_{\rm min}$, featuring an N shape with a local minimum at $(v_{\rm min}, \tau_{\rm min})$ (marked by the circle). For $\tau_{\rm d}\!>\!\tau_{\rm min}$ (e.g.~the horizontal dashed-dotted line), the equation $\tau_{\rm ss}(v)\!=\!\tau_{\rm d}$ features 3 solutions: the leftmost one (at extremely low slip velocities) is denoted by $v_{\rm stick}$ (square), the rightmost one is denoted by $v_{\rm ss}$ (triangle) and an intermediate one on the velocity-weakening branch (not marked). Two closed (homoclinic) orbits $(v(\xi),\tau(\xi))$, representing self-healing slip pulses, are added (dashed-dotted line for $\tau_{\rm d}/\tau_{\rm min}\!=\!1.05$ and the dashed line for $\tau_{\rm d}/\tau_{\rm min}\!=\!1.20$).}
\label{fig:fig1}
\end{figure}

The dimensionless frictional strength $f(v,\phi)\!\equiv\!\tau(v,\phi)/\sigma$, where $\sigma$ is the normal stress that presses the two bodies together, incorporates the $\phi$ dependence of the contact area and an additional logarithmic rheological dependence on $v$~\cite{Dieterich1979a,Ruina1983,Marone1998a,Nakatani2001,Baumberger2006Solid,Bar-Sinai2014}, and is presented in full detail in~\cite{SM}. Under steady sliding conditions, for which Eq.~\eqref{eq:dxi_phi} implies $\phi\=D/v$, the steady sliding friction curve $\tau_{\rm ss}(v)/\sigma\=f(v,\phi\=D/v)$ features an N shape, as demonstrated in numerous experiments~\cite{Bar-Sinai2014}. An example, to be used hereafter, is presented in Fig.~\ref{fig:fig1}. $\tau_{\rm ss}(v)$ is velocity-strengthening ($d\tau_{\rm ss}(v)/dv\!>\!0$) at extremely small slip velocities, then it becomes logarithmically velocity-weakening ($d\tau_{\rm ss}(v)/d\log(v)\=\hbox{const.}\!<\!0$) and eventually it becomes logarithmically velocity-strengthening, beyond the minimum of the curve at $(v_{\rm min}, \tau_{\rm min})$.

{\em A family of steady state pulse solutions}.---Eqs.~\eqref{eq:f_xi}-\eqref{eq:dxi_phi} correspond to two space dimensions, i.e.~translational invariance along the interface in the direction perpendicular to $x$ (the out-of-plane direction) is assumed. Here we focus on out-of-plane shear (mode-III symmetry), where the slip velocity $v(\xi)$ is perpendicular to the pulse propagation direction, and ${\cal F}_{_{\rm III}}(\beta)\=\sqrt{1-\beta^{2}}/(2\pi\beta c_{\rm s})$ (which vanishes as $\beta\!\to\!1$). The counterpart in-plane (mode-II) solutions readily follow, see~\cite{SM}.

The self-healing boundary conditions, previously expressed as $v(\xi\!\to\!\pm\infty)\!\to\!0$, take the form $v(\xi\!\to\!\pm\infty)\!=\!v_{\rm stick}$, where $v_{\rm stick}$ is an extremely low slip velocity that corresponds to the leftmost solution of $\tau_{\rm ss}(v)\=\tau_{\rm d}$ (cf.~Fig.~\ref{fig:fig1}). Consequently, a slip pulse corresponds to a closed (homoclinic) orbit in the $(v(\xi),\tau(\xi))$ plane, which starts and ends at $(v_{\rm stick},\tau_{\rm d})$. Two such closed orbits, for two different $\tau_{\rm d}$ values, are illustrated in Fig.~\ref{fig:fig1}.
\begin{figure}[ht!]
\centering
\includegraphics[width=0.48\textwidth]{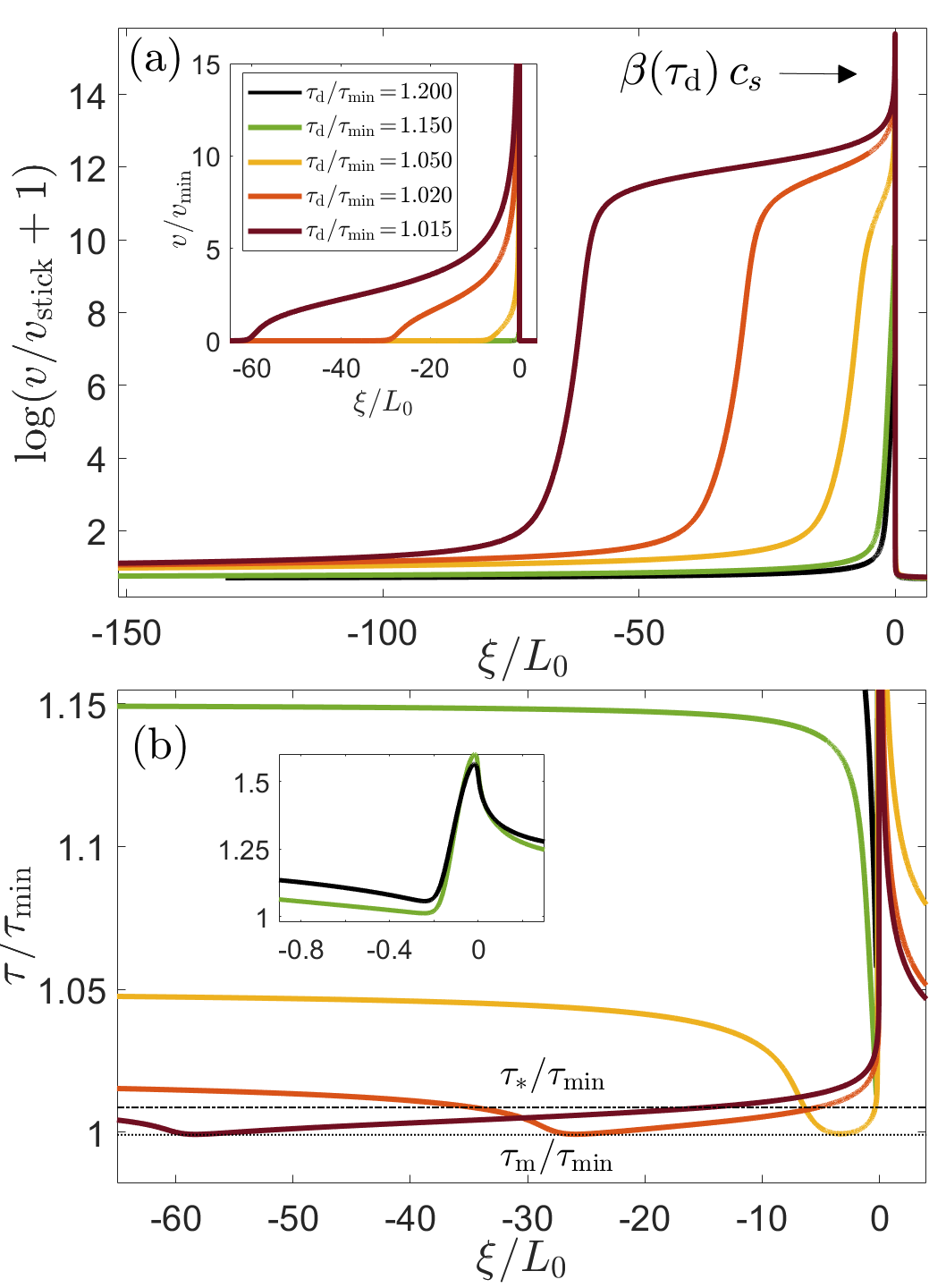}
\caption{(a) $\log(v/v_{\rm stick}+1)$ vs.~$\xi$ (normalized by the elasto-frictional length $L_0$, see~\cite{SM}) for several slip pulse solutions traveling from left to right at a velocity $\beta(\tau_{\rm d})c_{\rm s}$. The value of $\tau_{\rm d}$ for each curve is indicated in the legend in the inset, which presents the same results as the main panel, but in linear scale ($v$ is normalized here by $v_{\rm min}$). Note that the y-axis in the inset is truncated and that the curves for the two largest $\tau_{\rm d}$'s are not clearly discernible. (b) $\tau/\tau_{\rm min}$ vs.~$\xi/L_0$ for the same pulse solutions shown in panel (a). The minimal value $\tau_{\rm m}$ of the 3 lowest $\tau_{\rm d}$ curves (horizontal dotted line) and $\tau_*$ (horizontal dashed-dotted line, see Eqs.~\eqref{eq:scaling_beta}-\eqref{eq:scaling_L} and the discussion therein) are marked. (inset) A zoom in on the two largest $\tau_{\rm d}$ curves.}
\label{fig:fig2}
\end{figure}

We developed an accurate and robust numerical method to solve Eqs.~\eqref{eq:f_xi}-\eqref{eq:dxi_phi} with the self-healing boundary conditions $v(\xi\!\to\!\pm\infty)\!=\!v_{\rm stick}$, as detailed in~\cite{SM}. The formulated problem is shown to admit a family of steady state self-healing pulse solutions as a function of $\tau_{\rm d}$, a few of which are presented in Fig.~\ref{fig:fig2}. In Fig.~\ref{fig:fig2}a, $v(\xi)$ is presented, revealing a long healing tail at the trailing edge and a strong slip velocity amplification, by several orders of magnitude compared to the pulse's center, near the leading edge.

In Fig.~\ref{fig:fig2}b, the shear stress $\tau(\xi)$  --- which equals the frictional strength ---  is presented. $\tau(\xi)$ attains a peak near the leading edge, which is significantly larger than the driving stress $\tau_{\rm d}$, attained far ahead of the pulse (to the right, full relaxation is not shown. Note also that the peak value itself is truncated in the figure). It then attains a minimum value, which decreases with $\tau_{\rm d}$, but appears to converge to a value $\tau_{\rm m}$ that is close to $\tau_{\rm min}$. Finally, $\tau(\xi)$ slowly approaches $\tau_{\rm d}$ at the trailing edge, as the self-healing boundary condition is satisfied. The obtained solutions are highly accurate, featuring numerical convergence down to an error of ${\cal O}(10^{-15})$ and remarkable robustness with respect to the integration domain size~\cite{SM}.

{\em The pulse length and propagation velocity: theoretical considerations and scaling relations}.---We first focus on the variation of the pulse properties with $\tau_{\rm d}$, most notably the pulse length $L$ and the dimensionless pulse propagation velocity $\beta$. Note that the former does not appear in the problem formulation at all, but is rather defined a posteriori.

Under certain conditions, we expect $L$ --- to be operationally defined as the full width at half maximum of a logarithmic representation of $v(\xi)$ (see Fig.~\ref{fig:fig2}a and~\cite{SM} for full details) --- to diverge at a spacial driving stress $\tau_*$. The physics here is that a steady state pulse may sometimes be envisioned as composed of a crack-like rupture front --- yet another prominent spatiotemporal mode of rupture of frictional systems~\cite{Ben-Zion2001,Scholz2002,BarSinai2012,Svetlizky2019,Barras2019,Barras2020,lambert2021propagation} --- and a healing front, both propagating in the same direction at the same velocity~\cite{Brener2018}. $L$ is selected by the frictional interaction between these two fronts. Steady state crack-like fronts are related to the rightmost solution of $\tau_{\rm ss}(v)\=\tau_{\rm d}$, denoted by $v_{\rm ss}$ in~Fig.~\ref{fig:fig1}, where the stable fixed-point $v_{\rm ss}$ invades the $v_{\rm stick}$ fixed-point~\cite{BarSinai2012,Brener2018,Bar-Sinai2019}. A healing front corresponds to the opposite/inverse situation, where the nearly quiescent state $v_{\rm stick}$ invades the sliding one, $v_{\rm ss}$~\cite{Brener2018}.

If isolated steady state crack-like and healing fronts exist at the same propagation velocity at $\tau_{\rm d}\=\tau_*$, a pulse of infinite extent ($L\!\to\!\infty$) can be constructed by simply superimposing these two non-interacting fronts. This physical picture was validated for frictional systems of small height~\cite{Brener2018} and is expected to remain valid for the infinite height systems considered here. It is relevant for pulses that probe the velocity-strengthening branch of the friction curve, where $v_{\rm ss}$ resides. This is demonstrated by the closed $v\!-\!\tau$ orbit corresponding to $\tau_{\rm d}/\tau_{\rm min}\=1.05$ in~Fig.~\ref{fig:fig1}. Consequently, we expect the physical picture of pulses being viewed at interacting crack-like and healing fronts to be valid for $\tau_{\rm d}$ near $\tau_{\rm min}$, and $L$ to diverge at $\tau_{\rm d}\=\tau_*$, which is very close to $\tau_{\rm min}$ (below which $v_{\rm ss}$ does not exist anymore).

\begin{figure}[ht!]
\centering
\includegraphics[width=0.5\textwidth]{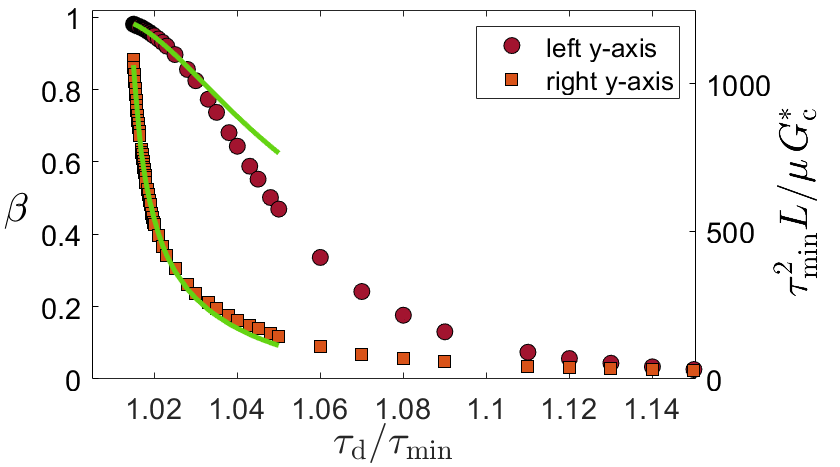}
\caption{$\beta(\tau_{\rm d})$ (left y-axis) and $L(\tau_{\rm d})$ (right y-axis, normalized by $\mu G^{*}_{\rm c}/\tau^2_{\rm min}$) of the obtained pulse solutions. We set $G^{*}_{\rm c}\!=\!0.65$ J/m$^2$, previously obtained for the corresponding crack-like rupture~\cite{Barras2020}. The solid lines correspond to the theoretical predictions in Eqs.~\eqref{eq:scaling_beta}-\eqref{eq:scaling_L}, see text for details.}
\label{fig:fig3}
\end{figure}

The existence of a minimum $(v_{\rm min}, \tau_{\rm min})$ of the friction curve and of $v_{\rm ss}$ are by no means necessary conditions for the existence of steady state pulses. The latter correspond to closed $v\!-\!\tau$ orbits that start and end at $(v_{\rm stick},\tau_{\rm d})$, and do not necessarily probe the minimum of the friction curve and the velocity-strengthening branch near it. This is expected to be the case for larger values of $\tau_{\rm d}$, away from the minimum $\tau_{\rm min}$, as is indeed demonstrated the closed $v-\tau$ orbit corresponding to $\tau_{\rm d}/\tau_{\rm min}\=1.2$ in~Fig.~\ref{fig:fig1}. In this regime, we expect $L$ to depend only mildly on $\tau_{\rm d}$. These expectations are verified in Fig.~\ref{fig:fig3} (right y-axis), where $L(\tau_{\rm d})$ is plotted and observed to strongly increase with decreasing $\tau_{\rm d}$, possibly consistent with a divergence as $\tau_{\rm min}$ is approached, and to vary mildly with $\tau_{\rm d}$ for larger values of $\tau_{\rm d}$. The corresponding results for $\beta(\tau_{\rm d})$ appear in Fig.~\ref{fig:fig3} (left y-axis), where $\beta$ is observed to become strongly inertial ($\beta\!\to\!1$) for small $\tau_{\rm d}$ and quasi-static for larger values.

To understand the behavior of $L(\tau_{\rm d})$ and $\beta(\tau_{\rm d})$ with decreasing $\tau_{\rm d}$, near $\tau_{\rm min}$, we first note that the rate dependence of $\tau_{\rm ss}(v)$ in Fig.~\ref{fig:fig1} is predominantly logarithmic, i.e.~rather weak. The recently developed theory of unconventional singularities of frictional rupture~\cite{brener2021unconventional,Brener2021JMPS} predicts that for weak rate dependence the order of the edge singularity experienced by various fields differs from the classical $-\tfrac{1}{2}$ singularity of fracture mechanics only mildly~\cite{brener2021unconventional}. With this in mind, we explore the possibility that {\em some physical quantities} approximately follow scaling relations inspired by classical pulse solutions featuring the classical $-\tfrac{1}{2}$ edge singularity.

Classical slip pulses with Coulomb (rate independent) friction~\cite{Freund1979} feature a length $L$, where inside the pulse $\tau(\xi)\=\tau_{\rm res}$ (the residual stress $\tau_{\rm res}$ corresponds to a dynamic/sliding friction coefficient of magnitude $\tau_{\rm res}/\sigma$) and out of it $v(\xi)\=0$. The transition from static Coulomb friction out of the pulse to dynamic/sliding Coulomb friction inside is characterized by a finite fracture energy $G_{\rm c}$, associated with a cohesive zone slip of magnitude $\delta_{\rm c}$~\cite{Palmer1973,Ida1972,Freund1979}. Formally, the boundary condition $\tau(\xi)\=\tau_{\rm res}$ inside the pulse is valid for slip $\delta$ satisfying $\delta\!>\!\delta_{\rm c}$ (i.e.~out of the cohesive zone). Solutions featuring a $-\tfrac{1}{2}$ power-law divergence near the leading edge at $\xi_{\rm p}$, and no divergence (but a discontinuous derivative) at the trailing edge, take the form $v(\xi)\=v_0\sqrt{(L+\xi_{\rm p}-\xi)/(\xi-\xi_{\rm p})}$~\cite{Freund1979}. The slip velocity at the middle of the pulse, $v_0$, satisfies $2c_{\rm s}\beta\=v_0 \mu \sqrt{1-\beta^2}\,(\tau_{\rm d}-\tau_{\rm res})^{-1}$ and $L$ satisfies $\pi L\=\mu G_{\rm c}\sqrt{1-\beta^2}\,(\tau_{\rm d}-\tau_{\rm res})^{-2}$~\cite{Freund1979}.

The above relations do {\em not} constitute a complete solution, as they feature three unknowns --- $\beta$, $L$ and $v_0$ --- and only two constraints. However, as our rate-and-state slip pulses solutions for $\tau_{\rm d}$ near $\tau_{\rm min}$ involve a characteristic slip velocity $v_{\rm min}$ (absent in the classical problem), we identify $v_0$ with $a_v v_{\rm min}$, $a_v$ being a dimensionless coefficient of ${\cal O}(1)$. Moreover, we identify the stress $\tau_{\rm res}$ with $\tau_*$ --- the hypothesized stress at which $L$ diverges ---, expected to be very close to $\tau_{\rm min}$. Finally, as we expect our pulses to feature an effective fracture energy $G_{\rm c}$ that is similar to their crack-like counterparts, we set $G_{\rm c}\=a_{_{\rm G}} G^{*}_{\rm c}$, with $a_{_{\rm G}}$ being a dimensionless coefficient of ${\cal O}(1)$ and $G^{*}_{\rm c}$ is the known crack-like rupture value~\cite{Barras2020}.

Taken together, we obtain
\begin{eqnarray}
\label{eq:scaling_beta}
\beta&=&a_v\left(\frac{v_{\rm min}\,\mu}{2\tau_{\rm min}\,c_{\rm s}}\right)\frac{\sqrt{1-\beta^2}}{(\tau_{\rm d}/\tau_{\rm min}-\tilde{\tau}_*)} \ ,\\
\label{eq:scaling_L}
\tilde{L}&\equiv&\frac{\tau_{\rm min}^2L}{\mu\,G^{*}_{\rm c}}=a_{_{\rm G}}\frac{\sqrt{1-\beta^2}}{\pi(\tau_{\rm d}/\tau_{\rm min}-\tilde{\tau}_*)^2}\ ,
\end{eqnarray}
where $\tilde{L}$ is the nondimensionalized $L$ and $\tilde{\tau}_*\!\equiv\!\tau_*/\tau_{\rm min}$. These predictions are quantitatively verified in Fig.~\ref{fig:fig3} (solid lines), with $\tilde\tau_*\!=\!1.0087$ ,$a_v\!=\!1.15$ and $a_{_{\rm G}}\!=\!0.69$ (note that $\tau_*$ is distinct from $\tau_{\rm m}$, cf.~Fig.~\ref{fig:fig2}b). We thus conclude that the approximate scaling relations in Eqs.~\eqref{eq:scaling_beta}-\eqref{eq:scaling_L}, together with the physical concepts and ideas incorporated into them, properly describe our steady state pulses for $\tau_{\rm d}$ near $\tau_*$.

{\em Energy dissipation and unconventional edge singularity}.---The above analysis indicates that our pulses for $\tau_{\rm d}$ close to $\tau_*$ (but not away from it) resemble in some respects classical pulses with Coulomb (rate independent) friction. Yet, on general physics grounds, we expect the intrinsic rate (and state) dependence of friction to make qualitative difference. To highlight this, we consider the energy budget associated with slip pulse propagation, in particular the local breakdown energy $\bar{G}(\delta)$ defined as~\cite{lambert2021propagation}
\begin{equation}
\label{eq:bar_G}
\bar{G}(\delta) = \int_0^\delta \left[\tau(\delta')-\tau_{\rm m} \right]d\delta' \ .
\end{equation}
Here $\delta(r)\=(\beta c_{\rm s})^{-1}\!\int_0^r\!\left[v(s)-v_{\rm stick}\right]ds$ ($s$ increases from the leading edge backwards into the pulse interior) is the slip accumulated by the pulse and $\tau_{\rm m}$ is the minimum of $\tau(\xi)$ for $\tau_{\rm d}$ close to $\tau_*$, marked in Fig.~\ref{fig:fig2}b.

\begin{figure}[ht!]
\centering
\includegraphics[width=0.45\textwidth]{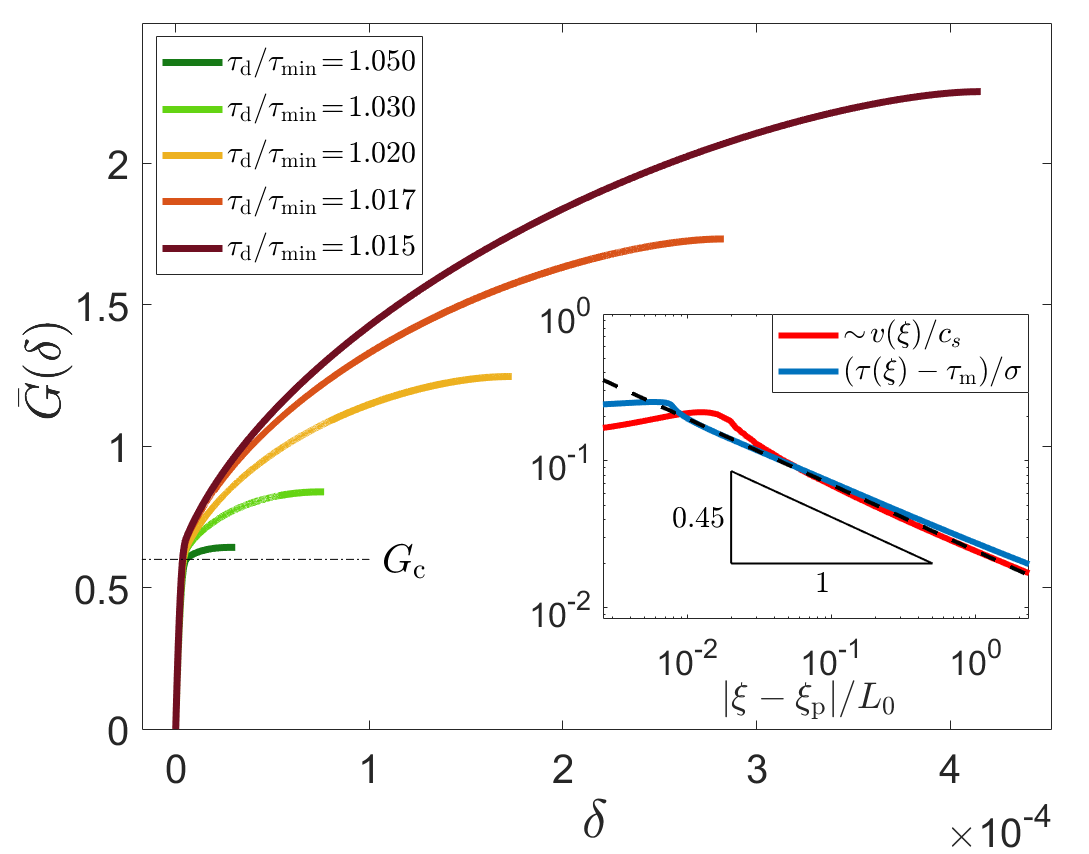}
\caption{$\bar{G}(\delta)$, see Eq.~\eqref{eq:bar_G}, for $\tau_{\rm d}$'s indicated in the legend. The different curves fan out at $G_{\rm c}\!\simeq\!0.6$ J/m$^2$ (horizontal dashed-dotted line). (inset) A singularity analysis of the near leading-edge fields, see legend and~\cite{SM}. The singularity order of the fields is $\simeq\!-0.45$ (dashed line and scaling triangle), see text for additional details.}
\label{fig:fig4}
\end{figure}

For classical pulses, we have $\tau_{\rm m}\=\tau_{\rm res}$ and $\bar{G}(\delta)\=G_{\rm c}$ for $\delta\!>\!\delta_{\rm c}$ (recall that for classical pulses $\tau(\delta)\=\tau_{\rm res}$ for $\delta\!>\!\delta_{\rm c}$), independently of the driving stress $\tau_{\rm d}$, and hence also of $\beta$ and $L$. In Fig.~\ref{fig:fig4}, we present $\bar{G}(\delta)$ of our solutions for $\tau_{\rm d}$ close to $\tau_*$. It is observed that all curves overlap at small $\delta$, indicating the existence of a well-defined effective fracture energy $G_{\rm c}$ (and of $\delta_{\rm c}$~\cite{SM}), corresponding to leading edge-localized dissipation. The various curves fan out at $G_{\rm c}\!\simeq\!0.6$ J/m$^2$, marked in Fig.~\ref{fig:fig4} (in quantitative agreement with the crack-like counterparts, see caption of Fig.~\ref{fig:fig3} and~\cite{Barras2020}). For larger $\delta$'s, the curves significantly deviate from $G_{\rm c}$ (reaching values as high as $\sim\!4G_{\rm c}$, attained for the smallest $\tau_{\rm d}$ considered), revealing excess dissipation that is distributed over the pulse length, which is entirely absent in classical pulses.

A recently developed theory~\cite{brener2021unconventional,Brener2021JMPS}, already mentioned above, showed that such spatially-extended excess dissipation is associated with the existence of unconventional singularities, i.e.~with near leading edge fields featuring a singularity order that differs from the classical $-\tfrac{1}{2}$ one. For the logarithmic rate dependence of the friction curve in Fig.~\ref{fig:fig1}, the singularity order deviation is predicted to be rather small, but the excess dissipation is large and increases with $L$. To test these predictions, we simultaneously fitted the slip velocity behind the pulse leading edge (i.e.~$\xi\!<\!\xi_{\rm p}$ in Fig.~\ref{fig:fig2}) to $v(\xi)\!\sim\!(\xi_{\rm p}\!-\!\xi)^\zeta$ and the shear stress ahead of the pulse leading edge (i.e.~$\xi\!>\!\xi_{\rm p}$ in Fig.~\ref{fig:fig2}) to $\tau(\xi)\!-\!\tau_{\rm m}\!\sim\!(\xi\!-\!\xi_{\rm p})^\zeta$, for the smallest $\tau_{\rm d}$ considered (largest $L$). The results are shown in the inset of Fig.~\ref{fig:fig4}, where the singularity order is $\zeta\!\simeq\!-0.45$ indeed deviates from $-\tfrac{1}{2}$, as predicted. These results clearly demonstrate that rate-and-state slip pulses reveal qualitative differences compared to their classical counterparts.

{\em Summary and outlook}.---In this work, we derived a family of steady state self-healing (solitonic) slip pulses in frictional systems for a realistic, experimentally supported, interfacial constitutive law. The physical properties of the emerging pulses have been thoroughly analyzed and theoretically explained. These results are of general importance for understanding spatiotemporal structures in driven nonlinear dissipative systems, featuring long-range interactions, and in particular for understanding the failure dynamics of frictional systems.

In the latter context, it is established that elasto-frictional instabilities --- where large amounts of stored elastic energy are abruptly released --- can spontaneously trigger long-lived pulse-like rupture (i.e.~propagating long distances without appreciably changing its properties). How do such dynamically generated long-lived pulse-like ruptures relate to the steady state pulse solutions derived here? To fully address this question, one should first determine the dynamic stability of the derived pulse solutions (when steady state conditions are not imposed), which is currently unknown. There are, however, some indications that these solutions might be dynamically unstable~\cite{Perrin1995,Brener2018,brantut2019stability}.

If true, then one can speculate that the growth rate of the dynamic instability is small and hence while steady state pulse solutions do not constitute a stable attractor (in the dynamical systems sense), a frictional system can nonetheless reside for rather long times near it. Addressing these important questions requires dynamical calculations that employ the obtained steady state solutions as initial conditions, which is a challenge for future investigations.

{\em Acknowledgements} This work has been supported by the Israel Science Foundation (grant no.~1085/20). E.B.~acknowledges support from the Ben May Center for Chemical Theory and Computation and the Harold Perlman Family.

\newpage
\onecolumngrid
\begin{center}
	\textbf{\large Supplemental Materials for:\\``Self-healing (solitonic) slip pulses in frictional systems''}
\end{center}

\setcounter{equation}{0}
\setcounter{figure}{0}
\setcounter{section}{0}
\setcounter{subsection}{0}
\setcounter{table}{0}
\setcounter{page}{1}
\makeatletter
\renewcommand{\theequation}{S\arabic{equation}}
\renewcommand{\thefigure}{S\arabic{figure}}
\renewcommand{\thesubsection}{S-\Roman{subsection}}
\renewcommand*{\thepage}{S\arabic{page}}
\twocolumngrid

The goal of this part is to provide additional technical details regarding the results reported on in the manuscript.

\subsection{The interfacial constitutive law}

We start by repeating Eqs.~\eqref{eq:f_xi}-\eqref{eq:dxi_phi} of the main text
\begin{eqnarray}
\label{eq:f_xi_SM}
\sigma f[v(\xi),\phi(\xi)]&=&\tau_{\rm d}-\mu\,{\cal F}(\beta)\!\int_{-\infty}^{\infty}\frac{v(z)}{z-\xi}dz \ ,\\
\label{eq:dxi_phi_SM}
-\beta\,c_{\rm s}\,\partial_{\xi}\phi(\xi)&=& g[v(\xi),\phi(\xi)] \ ,
\end{eqnarray}
where $\tau(v,\phi)\=\sigma f(v,\phi)$ was used in the former and the right hand side of the latter, $g(v,\phi)$, was left yet unspecified. Consequently, the interfacial constitutive law is described by two dimensionless functionals, $f(v,\phi)$ and $g(v,\phi)$, where the former may be viewed as a generalized ``friction coefficient'' (that is obviously not just a coefficient). In the rate-and-state friction framework, these take the form~\cite{Dieterich1979a,Ruina1983,Marone1998a,Nakatani2001,Baumberger2006Solid,Bar-Sinai2014,Brener2018}
\begin{eqnarray}
\label{eq:f_SM}
&& f(v,\phi)=\left[1+b\log(1+\phi/\phi _*)\right] \times \\
&&\qquad\qquad\qquad\qquad \left[\frac{f_0}{\sqrt{1+(v_*/v)^2}}+\alpha\log(1+|v|/v_*)\right]\ ,\nonumber\\
\label{eq:g_SM}
&&g(v,\phi)=1-\frac{|v|\phi}{D}\sqrt{1+(v_*/v)^2}\ .
\end{eqnarray}
Note that integral in Eq.~\eqref{eq:f_xi_SM}, which is nothing but the Hilbert transform of $v(\xi)$, is understood in the Cauchy principal value sense.

The physics behind Eqs.~\eqref{eq:f_SM}-\eqref{eq:g_SM}, and the corresponding experimental support, have been extensively discussed in previous literature~\cite{Dieterich1979a,Ruina1983,Marone1998a,Nakatani2001,Baumberger2006Solid,Bar-Sinai2014,Brener2018}. Here, we briefly note that $v_*$ is an extremely small slip velocity (cf.~Table~\ref{table:table1}) and that the function $\sqrt{1+(v_*/v)^2}$ that divides $f_0$ in Eq.~\eqref{eq:f_SM} ensures that $f(v,\phi)$ vanishes as $v\!\to\!0$, as required from general physics considerations~\cite{Estrin1996,Bar-Sinai2014}. The very same function also appears as a multiplicative factor in $g(v,\phi)$ in Eq.~\eqref{eq:dxi_phi_SM}, which ensures that for vanishingly small steady-state velocities, $\phi$ saturates after extremely long times to a finite value of $D/v_*$, rather than diverges. For practical purposes, though, this multiplicative factor makes a small quantitative difference. Hence, while it is included in the calculations, it is omitted from Eq.~\eqref{eq:dxi_phi} in the manuscript, where $g(v,\phi)\!\simeq\!1-|v|\phi/D$ was stated on the right hand side. Finally, the values of the parameters used in Eqs.~\eqref{eq:f_xi_SM}-\eqref{eq:g_SM} are given in Table~\ref{table:table1} and the emerging steady state friction curve is presented in Fig.~\ref{fig:fig1} in the manuscript.
\begin{table}[ht]
  \centering
  \begin{tabular}{|c|c|c|}
  \hline
  Parameter & Value & Units\\
  \hline
  $\mu$ & $9\!\times\!10^9$ & Pa\\ \hline
  $\sigma$ & $10^6$ & Pa\\ \hline
  $c_{\rm s}$ & $2739$ & m/s\\ \hline
  $D$ & $5\!\times\!10^{-7}$ &m \\ \hline
  $b$ & $0.075$ & -\\ \hline
  $v_*$ & $10^{-7}$ & m/s\\ \hline
  $f_0$ & $0.28$ & -\\ \hline
  $\phi_*$ & $3.3\!\times\!10^{-4}$ & s\\ \hline
  $\alpha$ & $0.005$ & -\\ \hline
\end{tabular}
\caption{Values of all parameters used (in MKS units).}
\label{table:table1}
\end{table}

\subsection{Numerical method and implementation}
\label{sec:numerical}

A major goal of this work is to solve Eqs.~\eqref{eq:f_xi_SM}-\eqref{eq:dxi_phi_SM}, together with Eqs.~\eqref{eq:f_SM}-\eqref{eq:g_SM}, for ${\cal F}(\beta)\={\cal F}_{_{\rm III}}(\beta)\=\sqrt{1-\beta^{2}}/(2\pi\beta c_{\rm s})$, where $\beta$ is understood as $\beta_{_{\rm III}}$, which is bounded by unity from above. The in-plane (mode-II symmetry) counterpart of this out-of-plane shear (mode-III symmetry) problem is discussed in Sect.~\ref{sec:mode-II}. The imposed self-healing boundary conditions, which define self-healing pulse solutions, take the form $v(\xi\!\to\!\pm\infty)\!=\!v_{\rm stick}$. We developed an accurate and robust numerical method to solve this problem, as we explain next.

We start by defining the following error function
\begin{equation}
\label{eq:error}
\Delta(\xi)\!\equiv\!\frac{\displaystyle\tau_{\rm d}\!-\!\left(\!\sigma f[v(\xi),\phi(\xi)] + \mu\,{\cal F}(\beta)\!\!\int_{-\infty}^{\infty}\!\!\frac{v(z)-v_{\rm stick}}{z-\xi}dz\!\right)}{\tau_{\rm d}} \ ,
\end{equation}
which identifies with Eq.~\eqref{eq:f_xi_SM} when the error vanishes everywhere in space, $\Delta(\xi)\=0$. Note that $v_{\rm stick}$ has been subtracted from the numerator of the integrand in Eq.~\eqref{eq:f_xi_SM}, which makes no difference as the Hilbert transform of a constant vanishes (yet, it is useful for the numerical procedure). Next, we represent the infinite domain $\xi\!\in\!(-\infty,\infty)$ by a finite, yet sufficiently large (a choice to be further discussed below), integration domain. The finite integration domain is discretized into $N$ spatial points $\xi_{i=1,2,...,N}$, forming a non-uniform grid. The selection of the non-uniform spatial distribution of the $N$ points will be discussed below. The basic fields in the problem, $\phi(\xi)$ and $v(\xi)$, are then discretized according to $\phi_i\!\equiv\!\phi(\xi_i)$ and $v_i\!\equiv\!v(\xi_i)$. Using the latter, we then discretize Eqs.~\eqref{eq:dxi_phi_SM} and~\eqref{eq:error}, where the discretized version of Eq.~\eqref{eq:error} results in $\Delta_i\!\equiv\!\Delta(\xi_i)$. Recalling that the dimensionless pulse propagation velocity $\beta$ is also unknown, we conclude that solving the problem amounts to finding $2N+1$ numbers $\{\phi_{i=1,2,...,N}, v_{i=1,2,...,N}, \beta\}$ that satisfy the discretized version of Eq.~\eqref{eq:dxi_phi_SM} and $\Delta_{i=1,2,...,N}\!\to\!0$ in the discretized version of Eq.~\eqref{eq:error}, together with the self-healing boundary conditions $v_1\=v_N\!\to\!v_{\rm stick}$.

To achieve this, we first set $v_1\=v_{\rm stick}+\delta{v}$ with $\delta{v}\!\ll\!v_{\rm stick}$, which ensures that the boundary condition on $v(\xi)$ is satisfied at one edge of the integration domain. Likewise, as $\phi(\xi)$ is in non-equilibrium steady state with $v(\xi)$ in this spatial regime (say, ahead of the pulse), we also set $\phi_1\=D/v_{\rm stick}-\delta\phi$ with $\delta\phi\!\ll\!D/v_{\rm stick}$. With $\phi_1$ at hand, we can readily obtain $\phi_{i=2,3,...,N}$ by integrating the discrete version of Eq.~\eqref{eq:dxi_phi_SM}, which is a first order differential equation in $\xi$ (it is solved by matrix inversion and requires the knowledge of $v_{i=1,2,...,N}$). It is important to note that due to the non-local (integro-differential) nature of the coupled Eqs.~\eqref{eq:f_xi_SM}-\eqref{eq:dxi_phi_SM}, $\{\phi_i\}$ implicitly depend on $v_{j=1,2,...,N}$ and $\beta$, so one should bear in mind that we actually obtain $\phi_i(\{v_j\},\beta)$ with $j\=1,2,...,N$. Next, we need to determine the remaining $N$ numbers $\{v_{i=2,3,...,N},\beta\}$, which is obviously related to requiring $\Delta_{i=1,2,...,N}\!\to\!0$. It is important to note that $\{\Delta_i\}$ implicitly depend on $v_{j=1,2,...,N}$, $\phi_{j=1,2,...,N}$ and $\beta$, i.e.~that we actually have $\Delta_i(\{v_k\},\phi_i(\{v_j\},\beta),\beta)$ with $k\=1,2,...,N$ and $j\=1,2,...,N$.

In principle, we could have used available nonlinear least-squares minimization algorithms to minimize $\Delta_i(\{v_k\},\phi_i(\{v_j\},\beta),\beta)$ for $i\=1,2,...,N$ with respect to $\{v_{i=2,3,...,N},\beta\}$ in order to obtain a solution. Yet, we found that analytically calculating the variation of $\Delta_i(\{v_k\},\phi_i(\{v_j\},\beta),\beta)$ with respect to $\{v_{i=2,3,...,N},\beta\}$, as a preparatory step for the least-squares minimization, greatly improves the efficiency, accuracy and robustness of the solution procedure. To this aim, we calculate the following derivatives
\begin{eqnarray}
\frac{d\Delta_i}{dv_j}&=&\frac{\pa\Delta_i}{\pa{v}_j} + \frac{\pa\Delta_i}{\pa\phi_k}\frac{\pa\phi_k}{\pa{v}_j} \ ,\\
\frac{d\Delta_i}{d\beta}&=&\frac{\pa\Delta_i}{\pa\beta} + \frac{\pa\Delta_i}{\pa\phi_k}\frac{\pa\phi_k}{\pa\beta}\ .
\end{eqnarray}
We then use $\Delta_i$, $d\Delta_i/dv_j$ and $d\Delta_i/d\beta$ (the latter two allow to construct the relevant Jacobian) as input to Matlab's nonlinear least-squares minimization function `lsqnonlin'~\cite{MATLAB}.

The procedure by which this numerical scheme is used to obtain slip pulse solutions --- including the choice of initial conditions for the minimization, the choice of the integration domain size and the choice of the non-uniform discrete grid --- are discussed in the next section. Before doing that, we note that the self-healing boundary condition at $i\=N$, $v_N\!\to\!v_{\rm stick}$, spontaneously emerges from the solution procedure (the degree to which it is satisfied is determined by the size of the integration domain, see below), but is not imposed. It is important to note in this context that the posed problem does not admit steady state crack-like rupture solutions, which correspond to the boundary condition $v_N\!\to\!v_{\rm ss}$ ($v_{\rm ss}$ is defined and discussed in the manuscript, and exists for $\tau_{\rm d}\!>\!\tau_{\rm min}$). Steady state crack-like rupture solutions exist for systems of finite height, say $H$, as was shown in~\cite{Brener2018,Bar-Sinai2019}, but not in the infinite system limit considered here, $H\!\to\!\infty$ (in the language of fracture mechanics, the latter leads to a diverging stress intensity factor~\cite{Broberg1999}). Finally, we note that as the Hilbert transform is formally divergent at the integration domain edges, $v(\xi)$ is slightly extrapolated beyond the integration domain such that the Cauchy principal value is well defined and convergent.

\subsection{Solution procedure and convergence}

The above-described numerical scheme is applied as follows. As we expect slip pulse solutions to feature a very strong amplification of the slip velocity $v(\xi)$ near the leading edge, the solution procedure should be able to take into account a very wide range of slip velocities (many orders of magnitude) and strong spatial localization. To address these physical features, we consider the field $\log[v(\xi)/v^*+1]$ (instead of $v(\xi)$ itself) and employ a non-uniform discretization grid. The spatial distribution of the grid points is selected such that it is dense where  $v(\xi)$ varies rapidly near the leading edge and more dilute elsewhere. In view of the non-uniform grid employed, the integral (Hilbert transform) in Eq.~\eqref{eq:error} is evaluated using a cubic interpolation of $v(\xi)$. Using conventional Hilbert transform routines that are based on a fast Fourier transform yields similar results, although undesired spurious oscillations typically emerge. As noted above, the discretized version of the state variable $\phi(\xi)$ is obtained by matrix inversion at each step, using Eq.~\eqref{eq:dxi_phi_SM} and the boundary condition for $\phi_1$.

The solution procedure requires the choice of suitable initial conditions, which take into account the expected existence of strong spatial localization and slip velocity amplification near the pulse leading edge and the long tail at the trailing edge. The solutions  discussed in this work are typically obtained by using a tilted triangle initial conditions for $\log[v(\xi)/v^*+1]$. The precise details of the tilted triangle initial conditions do not affect the resulting solution. Other initial conditions, such as a skewed Gaussian or a narrow rectangle, also lead to the same solution, albeit they require more steps to converge and additional heuristic strategies for meeting a given convergence level. We used  Matlab's nonlinear least-squares minimization function `lsqnonlin'~\cite{MATLAB}, as explained above, until $\max[\Delta_i]\!\sim\!10^{-15}$. That is, our convergence criterion is very strict, corresponding to an error of ${\cal O}(10^{-15})$ across the entire integration domain.

We found that obtaining solutions for relatively large values of $\tau_{\rm d}$, away from the minimum of the steady state friction curve, is generally easier. Consequently, we first obtained a pulse solution for large $\tau_{\rm d}$ and then reduced $\tau_{\rm d}$ by small increments, typically of magnitude $\tau_{\rm d}/\tau_{\rm min}\!\sim\!10^{-2}$. For each $\tau_{\rm d}/\tau_{\rm min}$ value, we used the converged pulse solution at the previous value as the initial condition. Finally, as the length of the pulse increases with decreasing $\tau_{\rm d}/\tau_{\rm min}$, the integration domain has been extended progressively and the grid was re-meshed accordingly, in order to account for the resulting variation in the position of the leading and trailing edges. To obtain the initial conditions for the extended integration domain, the field $v(\xi)$ of the previous solution was linearly interpolated inside the pulse and linearly extrapolated out of it, into the extended domain.
\begin{figure}[ht!]
\centering
\includegraphics[width=0.45\textwidth]{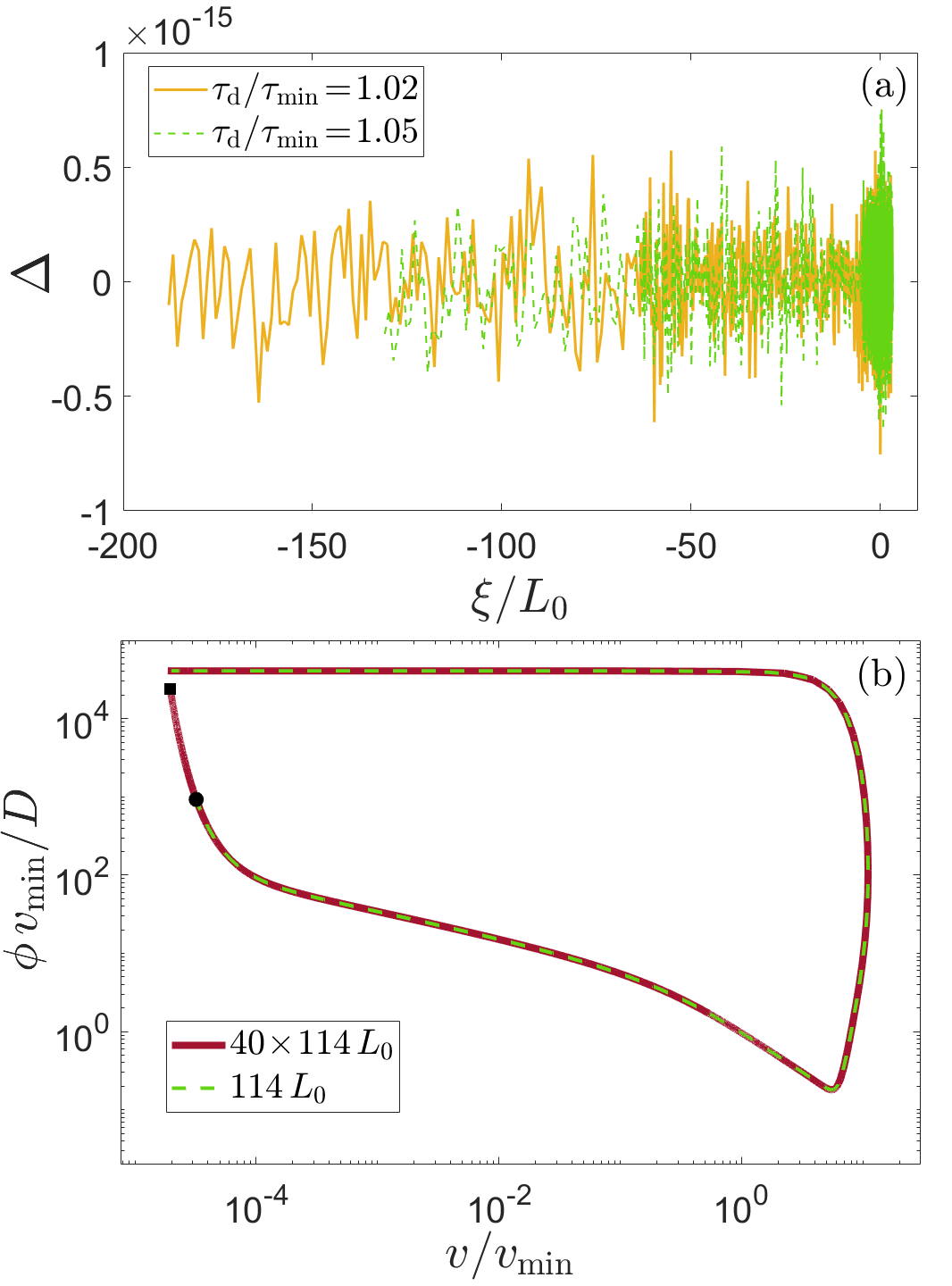}
\caption{(a) The error $\Delta(\xi)$, defined in Eq.~\eqref{eq:error}, for two pulse solutions (see legend). Note the $10^{-15}$ scale of the y-axis. (b) Two pulse solutions for the very same parameters (here $\tau_{\rm d}\!=\!1.050\tau_{\rm min}$), except for the size of the integration domain used (one is 40 times larger than the other, see legend), represented in the $v/v_{\rm min}\!-\!\phi v_{\min}/D$ plane. It is observed that the two solutions perfectly overlap over the shorter domain (up to the marked circle) and that the extended integration domain leads to a better resolution of the long healing tail as a closed (homoclinic) orbit is approached (the square).}
\label{fig:convergence}
\end{figure}

Using the above procedure, we derived slip pulse solutions over a wide range of driving stresses, ranging from $\tau_{\rm d}\=1.015 \tau_{\rm min}$ --- close to the minimum of steady state friction curve --- to $\tau_{\rm d}\=1.200 \tau_{\rm min}$ --- far away from it. Examples for solutions in this entire range are presented in Fig.~\ref{fig:fig2} in the manuscript, showing both the slip velocity field $v(\xi)$ and the shear stress field $\tau(\xi)$. Throughout the manuscript, we normalize the co-moving coordinate $\xi$ by the elasto-frictional length defined here as $L_0\!=\!\frac{\pi \mu D}{\sigma (f_0 b - \alpha)}$~\cite{Bar-Sinai2019}, in terms of the frictional parameters $D$, $f_0$, $b$ and $\alpha$, the shear modulus $\mu$ and the normal stress $\sigma$ (see Eqs.~\eqref{eq:f_xi_SM}-\eqref{eq:g_SM} and Table~\ref{table:table1}).

The obtained solutions are very accurate, featuring an error --- quantified by $\Delta(\xi)$ of Eq.~\eqref{eq:error} --- that is bounded from above by $10^{-15}$ (as already noted above). Two examples of $\Delta(\xi)$ are presented in Fig.~\ref{fig:convergence}a. In addition, the solutions are very robust against variations of the size of the integration domain. In Fig.~\ref{fig:convergence}b, we present two solutions in the $\phi\!-\!v$ plane for the very same set of parameters, where a different integration domain size has been used. It is observed that the two solutions perfectly overlap over the shorter domain (up to the marked circle) and that the extended integration domain leads to a better resolution of the long healing tail as a closed (homoclinic) orbit is approached.

The derived slip pulse solutions are characterized by their dimensionless propagation velocity $\beta(\tau_{\rm d})$, which is explicitly selected by the solution, and by their length $L(\tau_{\rm d})$, which does not appear in the problem formulation and hence is defined a posteriori. We operationally define $L$ as the full width at half maximum of $\log[v(\xi)/v_*+1]$. Note that the latter is similar to $\log[v(\xi)/v_{\rm stick}+1]$, which is presented in Fig.~\ref{fig:fig2}a in the manuscript, where both $v_*$ and $v_{\rm stick}$ are extremely small slip velocities. The difference is that $v_*$ is an interfacial parameter (that roughly determines the location of the local maximum of the steady state friction curve, cf.~Fig.~\ref{fig:fig1} in the manuscript) and $v_{\rm stick}$ somewhat varies with $\tau_{\rm d}$. Other operational definitions of $L$ give rise to quantitatively similar results. $\beta(\tau_{\rm d})$ and $L(\tau_{\rm d})$ are presented in Fig.~\ref{fig:fig3} in the manuscript. Note that $L$ is not normalized by $L_0$, but rather by $\mu\,G^{*}_{\rm c}/\tau_{\rm min}^2$, which is a more natural lengthscale in this context (cf.~Eq.~(4) in the manuscript).

It is important to note that our formulation differs from previous works aiming at solving related steady state pulse problems (e.g.~for different interfacial constitutive relations~\cite{Perrin1995,Garagash2012,brantut2019stability}), where the self-healing boundary conditions are a priori defined on a finite domain of size $L$. In such a formulation, the slip velocity is assumed to identically vanish outside the pulse, $v_{\rm stick}\!\to\!0$, and $L$ is self-consistently selected. Such boundary conditions imply that the resulting $v(\xi)$ features a discontinuous derivative at the pulse trailing edge and that the shear stress out of the pulse does not equal the frictional strength $\tau(v,\phi)$. That is, $\tau(v,\phi)$ in such formulations plays a dual role, i.e.~inside the pulse it determines the shear stress, while out of it $\tau(v,\phi)$ serves as an upper bound on the shear stress.

Mathematically speaking, the shear stress in such formulations is set equal to the frictional strength inside the pulse, but satisfies an inequality out of it, yet involving the very same $\tau(v,\phi)$ function. These features also imply that the long healing tail is not fully resolved in such solutions. Our formulation does not involve any of these simplifying assumptions, but rather requires that $\tau(v,\phi)$ determines the shear stress over the entire interface and hence fully resolves the long healing tail.

\subsection{The in-plane shear (mode-II symmetry) solutions}
\label{sec:mode-II}

The problem defined in Eqs.~\eqref{eq:f_xi_SM}-\eqref{eq:dxi_phi_SM} is two-dimensional, i.e.~translational invariance along the interface in the direction perpendicular to $\xi$ (the out-of-plane direction) is assumed. Equation~\eqref{eq:f_xi_SM} is valid for both out-of-plane shear (the so-called mode-III symmetry), where the slip velocity $v(\xi)$ is perpendicular to the pulse propagation direction, and in-plane shear (mode-II symmetry), where $v(\xi)$ is parallel to it, depending on the choice of the dimensionless function ${\cal F}(\beta)$. In the manuscript, we focus on mode-III symmetry, where ${\cal F}(\beta)\={\cal F}_{_{\rm III}}(\beta_{_{\rm III}})$ (see Sect.~\ref{sec:numerical} above). Here, $\beta$ is understood to stand for $\beta_{_{\rm III}}$ and ${\cal F}_{_{\rm III}}(\beta_{_{\rm III}})$ vanishes as $\beta_{_{\rm III}}\!\to\!1$.

\begin{figure}[ht!]
\centering
\includegraphics[width=0.45\textwidth]{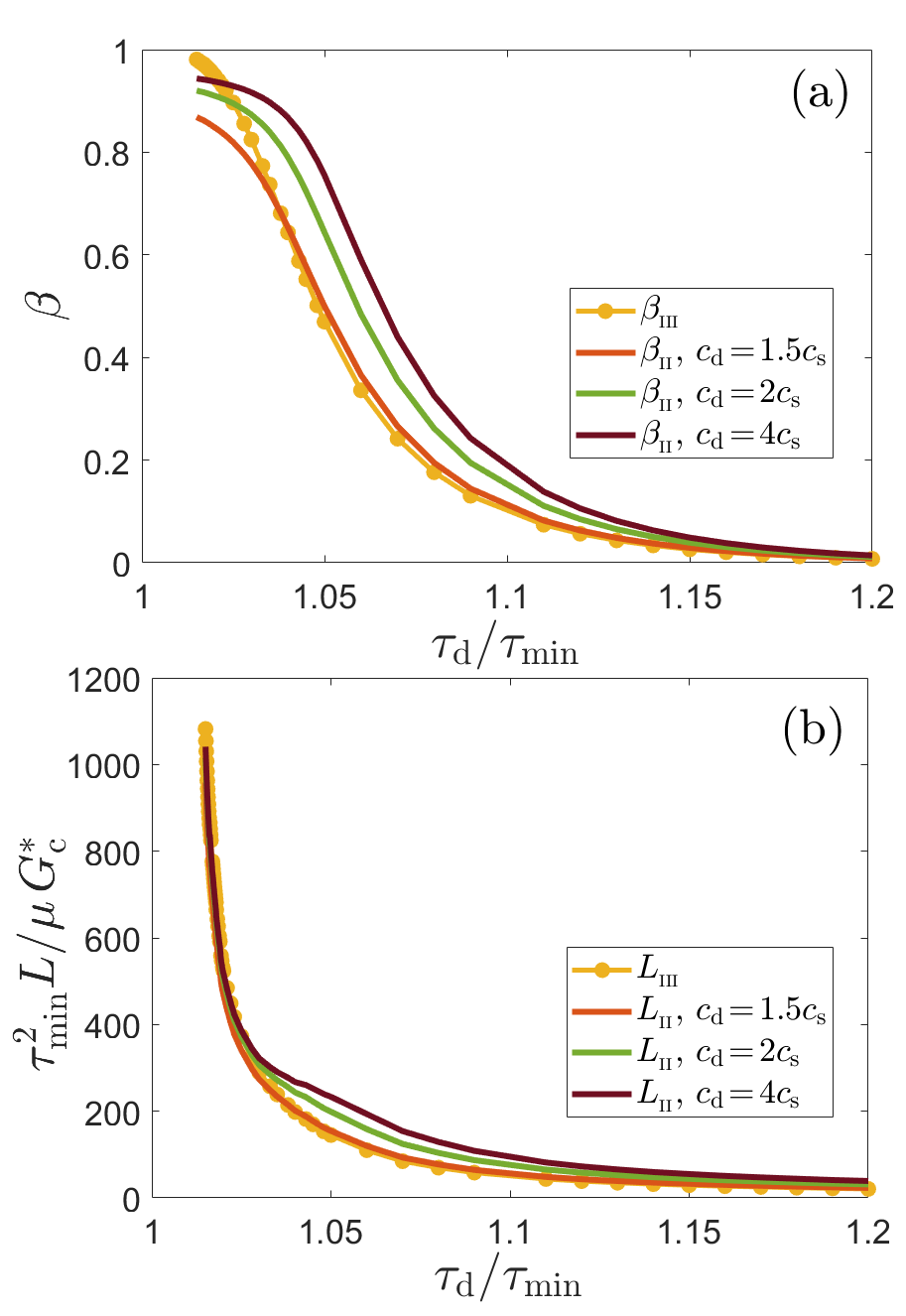}
\caption{(a) The dimensionless pulse propagation velocity $\beta$ vs.~$\tau_{\rm d}/\tau_{\rm min}$ for the mode-II solutions (marked as $\beta_{_{\rm II}}$ in the legend) for 3 values of $c_{\rm d}/c_{\rm s}$ (see legend). The corresponding results for mode-III (marked as $\beta_{_{\rm III}}$ in the legend), already reported on in Fig.~\ref{fig:fig3} in the manuscript, are added for comparison. The mode-II and mode-III results are similar, except for the limiting propagation velocity, which is $c_{_{\rm R}}\!<\!c_{\rm s}$ for the former ($c_{_{\rm R}}$ depends on $c_{\rm d}/c_{\rm s}$) and $c_{\rm s}$ for the latter. (b) The same as panel (a), but for the (normalized) pulse length $L$.}
\label{fig:mode-II}
\end{figure}

For in-plane shear (mode-II symmetry), we have
\begin{equation}
{\cal F}_{_{\rm II}}(\beta_{_{\rm II}})=\frac{D(\beta_{_{\rm II}})}{2\pi \beta_{_{\rm II}}^3 \sqrt{1-\beta_{_{\rm II}}^2}\,c_{\rm s}} \ ,
\end{equation}
where $D(\beta)\=4\sqrt{1-\beta^2}\sqrt{1-\beta^2(c_{\rm s}/c_{\rm d})^2}-(2-\beta^2)^2$ is the Rayleigh function (that vanishes at the Rayleigh wave-speed $c_{_{\rm R}}$) and $c_{\rm d}$ is the dilatational wave-speed. Once the results for mode-III are available, the corresponding results for mode-II are readily obtained by a simple transformation (note that the mode-II problem involves another material parameter, $c_{\rm d}/c_{\rm s}\!>\!1$, and that we focus on sub-Rayleigh solutions, $c_{\rm p}\!<\!c_{_{\rm R}}$). To see this, we note that if $\beta_{_{\rm III}}$, $v(\xi_{_{\rm III}})$ and $\phi(\xi_{_{\rm III}})$ are known, then $\beta_{_{\rm II}}$ is obtained by solving ${\cal F}_{_{\rm II}}(\beta_{_{\rm II}})\={\cal F}_{_{\rm III}}(\beta_{_{\rm III}})$, and $v(\xi_{_{\rm II}})$ and $\phi(\xi_{_{\rm II}})$ are obtained by the following coordinate transformation $\xi_{_{\rm III}}\to \xi_{_{\rm II}} \beta_{_{\rm III}}/\beta_{_{\rm II}}$. The mode-II results, emerging from this procedure, are reported in Fig.~\ref{fig:mode-II} (for a few $c_{\rm d}/c_{\rm s}$ values) and shown to be similar to the corresponding mode-III results (except that their limiting velocity $c_{_{\rm R}}$, which slightly depends on $c_{\rm d}/c_{\rm s}$, is smaller than $c_{\rm s}$).

\subsection{Additional technical details and\\ supporting results}

\hspace{-0.4cm}{\em Determining the unconventional singularity order}.---The near leading-edge singularity analysis for the $\tau_{\rm d}\=1.015 \tau_{\rm min}$ pulse solution is presented in the inset of Fig.~\ref{fig:fig4} in the manuscript. The analysis follows a related procedure employed in~\cite{Barras2020} for crack-like frictional rupture. That is, we simultaneously fit the slip velocity behind the pulse leading edge (i.e.~$\xi\!<\!\xi_{\rm p}$ in Fig.~\ref{fig:fig2} in the manuscript) to $v(\xi)/c_{\rm s}\!\sim\!(\xi_{\rm p}-\xi)^\zeta$ and the shear stress ahead of the pulse leading edge (i.e.~$\xi\!>\!\xi_{\rm p}$ in Fig.~\ref{fig:fig2} in the manuscript) to $(\tau(\xi)-\tau_{\rm m})/\sigma\!\sim\!(\xi-\xi_{\rm p})^\zeta$. Note that for presentational purposes, $v(\xi)/c_{\rm s}$ was multiplied by a dimensionless factor in order for the two fields to appear on the same scale in the inset of Fig.~\ref{fig:fig4} in the manuscript. The two fitting parameters are the singularity order $\zeta$ and the effective edge location $\xi_{\rm p}$, which is tightly constrained to reside between the maxima of the two fields. The best fit estimate of the singularity order is $\zeta\!\simeq\!-0.45$, as shown in the manuscript.

The recently developed theory of unconventional singularities of frictional rupture~\cite{brener2021unconventional,Brener2021JMPS} predicts that the degree of deviation from the classical $-\tfrac{1}{2}$ singularity order depends on the rate dependence of the frictional strength, i.e.~on the magnitude of $d\tau_{\rm ss}/dv$, and on the rupture velocity $\beta$. Moreover, for cases in which the near edge frictional strength is rate-strengthening, $d\tau_{\rm ss}/dv\!>\!0$, as is the case in our solutions near the minimum of the steady state friction curve (cf.~the dashed-dotted closed orbit in Fig.~\ref{fig:fig1} in the manuscript, which corresponds to $\tau_{\rm d}\=1.050 \tau_{\rm min}$ and which extends beyond the minimum of the steady state friction curve), the theory predicts that the unconventional singularity order would be smaller (in absolute value) than $-\tfrac{1}{2}$. The obtained value, $\zeta\!\simeq\!-0.45$, shows that our pulse solutions indeed feature an unconventional edge singularity and that indeed its order is smaller in absolute value compared to the conventional square root singularity, as predicted theoretically.

Finally, the unconventional singularities theory of frictional rupture~\cite{brener2021unconventional,Brener2021JMPS} predicts that the spatially-extended excess dissipation $\bar{G}(\delta)/G_{\rm c}$ --- defined in Eq.~\eqref{eq:bar_G} in the manuscript --- increases with both the propagation velocity $\beta$, as it leads to larger deviations from the $-\tfrac{1}{2}$ singularity, and with the pulse length $L$, as dissipation is accumulated over a more extended spatial region~\cite{brener2021unconventional,Brener2021JMPS}. This prediction is also consistent with the results of Fig.~\ref{fig:fig4} in the manuscript (main panel), as both $\beta$ and $L$ increase with decreasing $\tau_{\rm d}$.\\

\hspace{-0.4cm}{\em The field $\phi(\xi)$ in the pulse interior}.---The pulse fields $v(\xi)$ and $\tau(\xi)$ are presented in Fig.~\ref{fig:fig2} in the manuscript, although the two basic fields in the problem formulated in Eqs.~\eqref{eq:f_xi_SM}-\eqref{eq:dxi_phi_SM} are $v(\xi)$ and $\phi(\xi)$. Some information about $\phi(\xi)$ is provided in the $\phi\!-\!v$ plot in Fig.~\ref{fig:convergence}b, yet it is not spatial information per se. To gain insight into the latter, we consider here $\log(1-g[v(\xi),\phi(\xi)])$. This quantity is of interest as it provides a measure for the degree by which the interfacial internal state field $\phi(\xi)$ reaches a non-equilibrium steady state with the slip velocity field $v(\xi)$. The latter is attained when $g(v,\phi)\=0$ on the right hand side of Eq.~\eqref{eq:dxi_phi_SM}, implying that $\phi(\xi)\!\simeq\!D/v(\xi)$. This corresponds to $\log(1-g[v(\xi),\phi(\xi)])\=0$. Consequently, the deviation of $\log(1-g[v(\xi),\phi(\xi)])$ from zero quantifies the degree by which $\phi(\xi)$ is out of steady state with $v(\xi)$.

\begin{figure}[ht!]
\centering
\includegraphics[width=0.49\textwidth]{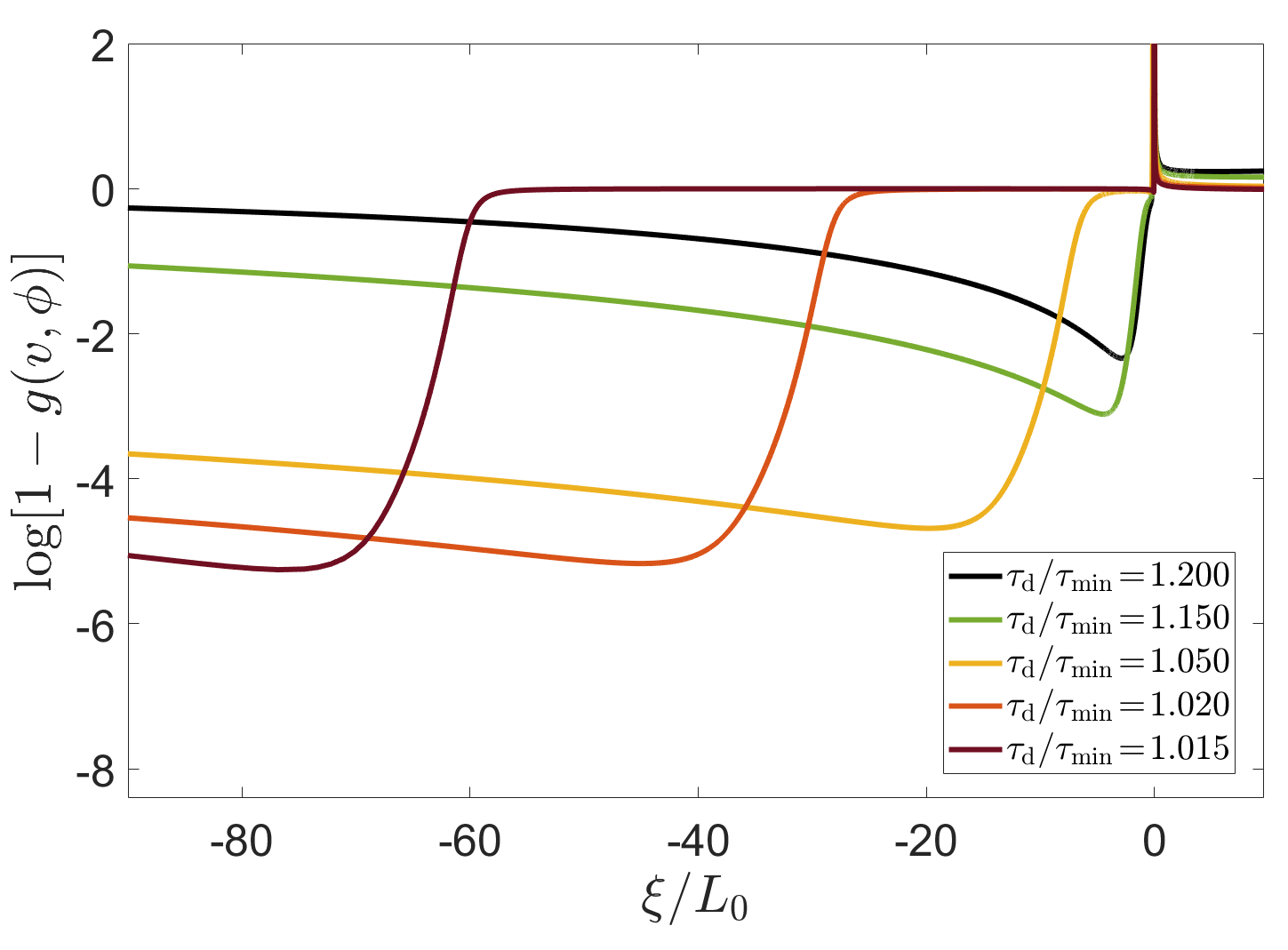}
\caption{$\log(1-g[v(\xi),\phi(\xi)])$ vs.~$\xi/L_0$ for the $\tau_{\rm d}$ values indicated in the legend, see text for discussion.}
\label{fig:1-g}
\end{figure}

In Fig.~\ref{fig:1-g}, we present $\log(1-g[v(\xi),\phi(\xi)])$ for several $\tau_{\rm d}$ values, spanning the entire range available. It is observed that for $\tau_{\rm d}$ close to $\tau_{\rm min}$, $\phi(\xi)$ reaches a non-equilibrium steady state with $v(\xi)$. The latter corresponds to $\log(1-g[v(\xi),\phi(\xi)])\=0$, which also clearly reveals the slip pulse size $L$. The very same physics is revealed by the $\tau_{\rm d}\=1.015 \tau_{\rm min}$ closed orbit in the $v\!-\!\tau$ plane, shown in Fig.~\ref{fig:fig1} in the manuscript, where the solution overlaps the steady state friction curve near its minimum (and above it). The situation is qualitatively different for the two larger $\tau_{\rm d}$ curves in Fig.~\ref{fig:1-g}, for which $\phi(\xi)$ does not reach a non-equilibrium steady state with $v(\xi)$. This situation is also evident in the $\tau_{\rm d}\=1.20 \tau_{\rm min}$ closed orbit in the $v\!-\!\tau$ plane, shown in Fig.~\ref{fig:fig1} in the manuscript. This change of character of the pulse solutions depending on the proximity of $\tau_{\rm d}$ to $\tau_{\rm min}$ has been highlighted in the manuscript (in relation to the scaling theory in Eqs.~\eqref{eq:scaling_beta}-\eqref{eq:scaling_L} therein and the connection --- and sometimes lack of --- between steady state pulse solutions and steady state crack-like/healing front, see manuscript for discussion).

\hspace{-0.4cm}{\em The fracture energy $G_{\rm c}$ and cohesive zone slip $\delta_{\rm c}$}.---The existence of an effective fracture energy $G_{\rm c}$ of the obtained pulses, corresponding to the leading edge dissipation, is indicated by the small $\delta$ overlap of the different $\bar{G}(\delta)$ curves shown in Fig.~\ref{fig:fig4} in the manuscript. The latter also implies the existence of a cohesive zone slip $\delta_{\rm c}$ over which the dissipation quantified by $G_{\rm c}$ occurs, defined as $\bar{G}(\delta_{\rm c})\=G_{\rm c}$. However, $\delta_{\rm c}$ is too small to be discernible on Fig.~\ref{fig:fig4} in the manuscript. Our goal here is to elucidate this point.

The leading edge-localized dissipation $G_{\rm c}$ is related to a strong strength reduction near the pulse leading edge, over which a slip of magnitude $\delta_{\rm c}$ is accumulated. This strong frictional strength reduction is associated in the rate-and-state constitutive framework with the evolution of the internal state field $\phi$. It has been shown~\cite{Cocco2002,Bizzarri2003} that while the rate-and-state constitutive framework does not make explicit reference to the slip $\delta$, the strength reduction from the peak stress (say $\tau_0$) --- reached after the very initial increase in slip velocity near the pulse leading edge --- to a level $\tau_{\rm c}$ at $\delta\=\delta_{\rm c}$ (where $\tau_{\rm c}$ is still larger than $\tau_{\rm m}$) follows an effective linear slip-weakening law of the form $\tau(\delta)\!\simeq\!\tau_0-(\tau_0-\tau_{\rm c})\delta/\delta_{\rm c}$.

\begin{figure}[ht!]
\centering
\includegraphics[width=0.49\textwidth]{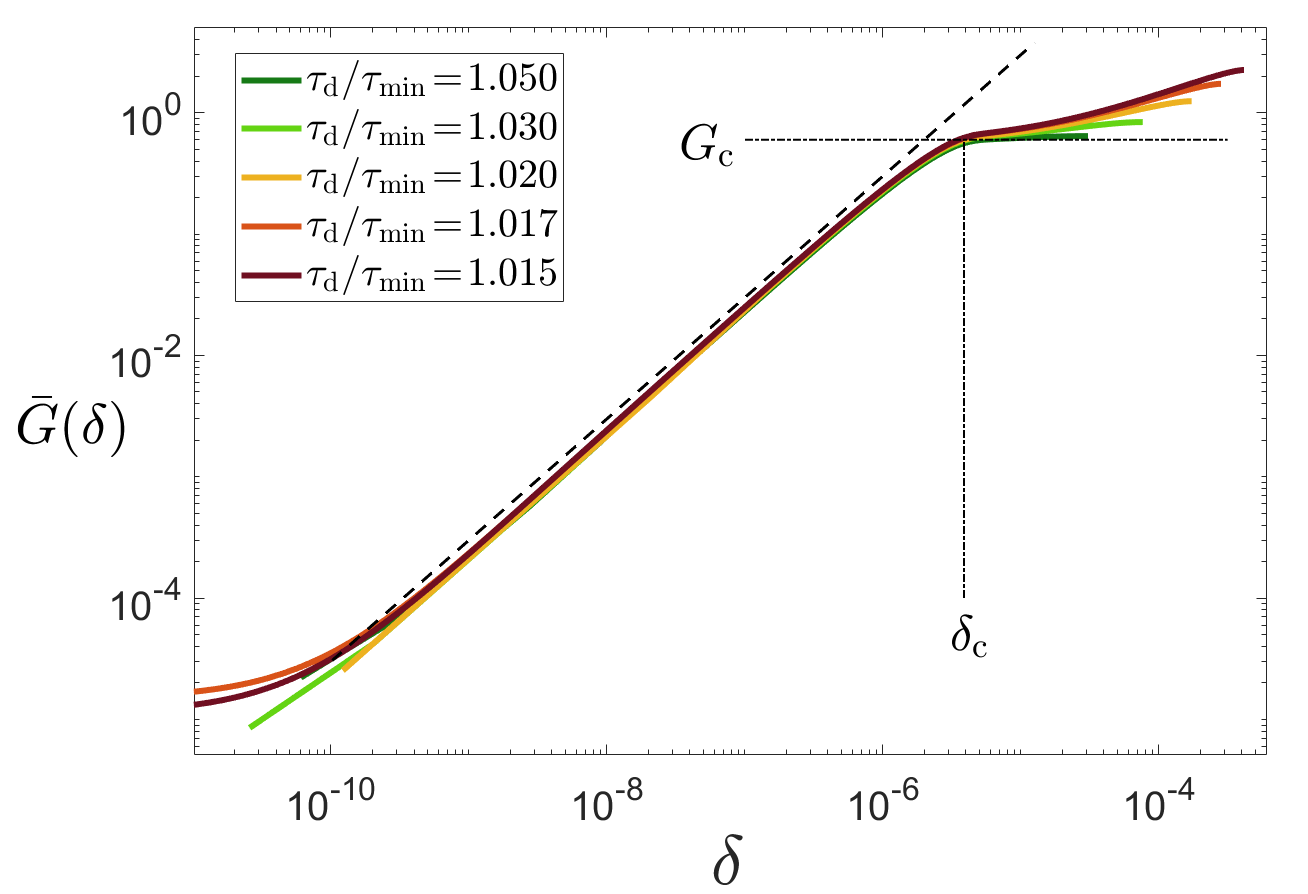}
\caption{The same as Fig.~\ref{fig:fig4} in the manuscript, except that here $\bar{G}(\delta)$ is presented on a double-logarithmic scale. The dashed line corresponds to $\bar{G}(\delta)\!\propto\!\delta$, as predicted for $\delta\!<\!\delta_{\rm c}$. $\bar{G}(\delta_{\rm c})\!=\!G_{\rm c}$ is represented by the dashed-dotted lines ($G_{\rm c}\!=\!0.6$ J/m$^2$ by the horizontal one, also marked in Fig.~\ref{fig:fig4} in the manuscript, and $\delta_{\rm c}$ by the vertical one).}
\label{fig:cohesive}
\end{figure}

Using the latter relation inside the definition of $\bar{G}(\delta)$ in Eq.~5 in the manuscript, we obtain $\bar{G}(\delta)\!\propto\!\delta$ in this small $\delta$ regime to leading order. In Fig.~\ref{fig:cohesive}, we replot the results for $\bar{G}(\delta)$ presented in the main panel of Fig.~\ref{fig:fig4} in the manuscript, this time on a double-logarithmic scale. $\bar{G}(\delta)\!\propto\!\delta$ for $\delta\!<\!\delta_{\rm c}$ is evident, as predicted, where $\delta_{\rm c}$ is clearly discernible now.

\hspace{-0.4cm}{\em Existence of other families of solutions}.---In this work, we exclusively focused on one family of steady state pulse solutions, where the possibility that other solutions exist as well has not been discussed. We do not claim that such solutions do not exist, and in fact while deriving the presented family of solutions, we found clear indications that additional solutions do exist. These additional pulse solutions have not been systematically explored and thoroughly analyzed so far; yet, for a given driving stress $\tau_{\rm d}$, all of the additional pulse solutions we managed to trace featured a larger size compared to $L(\tau_{\rm d})$ presented in Fig.~\ref{fig:fig3} in the main text. Hence, while we do not have a strict proof, we strongly suspect that the family of solutions discussed in this work feature the smallest size $L$ for a given $\tau_{\rm d}$. Future work should address additional steady state pulse solutions.


%

\end{document}